\documentclass{ceurart}

\usepackage{algorithm}
\usepackage{algorithmic}
\usepackage{tikz}
\usepackage{pgfplots}
\usepackage[mode=buildnew]{standalone}
\pgfplotsset{compat=1.18}
\usepackage{gensymb}
\usepackage{booktabs}
\usetikzlibrary{fit,arrows,positioning,arrows.meta}
\usetikzlibrary{patterns}
\usetikzlibrary{shapes.arrows}
\usepackage{graphicx}
\usepackage{textcomp}
\usepackage{geometry}
\begin{document}

\title{Angle of Arrival Estimation Using SRS in 5G NR Uplink Scenarios}

\author[1,2]{Thodoris Spanos}[%
orcid=0000-0002-0877-7063,
email=tspanos@ece.upatras.gr]
\author[3]{Fran Fabra}[%
orcid=0000-0001-8100-1520,
email=franciscojose.fabra@uab.cat]

\author[3]{Jos\'e A. L\'opez-Salcedo}[%
orcid=0000-0002-5392-6005,
email=jose.salcedo@uab.cat
]
\author[3]{Gonzalo Seco-Granados}[%
orcid=0000-0003-2494-6872,
email=gonzalo.seco@uab.cat
]
\author[2]{Nikos Kanistras}[%
orcid=0000-0002-7337-1107,
email=nikos.kanistras@loctio.com
]
\author[4]{Ivan Lapin}[%
orcid=0000-0002-1847-5499,
email=ivan.lapin@esa.int
]
\author[1]{Vassilis Paliouras}[%
orcid=0000-0002-1414-7500,
email=paliuras@ece.upatras.gr
]
\address[1]{Dept. of Electrical and Computer Engineering, University of Patras, Greece}
\address[2]{Loctio, Greece}
\address[3]{Dept. of Telecommunication and Systems Eng., Universitat Aut\`onoma de Barcelona (UAB), Spain}
\address[4]{Radio Navigation Systems and Techniques Section, European Space Agency, The Netherlands}
\copyrightclause{Copyright for this paper by its authors.\\Use permitted under the Creative Commons License Attribution 4.0 International (CC BY 4.0).}

\conference{WIPHAL 2024: Work-in-Progress in Hardware and Software for
Location Computation, June 25-27, 2024, Antwerp, Belgium}
\maketitle

\begin{abstract}
This paper presents a comprehensive exploration of Angle of Arrival (AoA) estimation techniques in 5G environments, using the Sounding Reference Signal (SRS) in Uplink scenarios both in simulations and with actual measurements. Leveraging 5G capabilities, we investigate AoA algorithms for single-base station positioning. The study includes simulations and practical tests on a developed dedicated testbed featuring a base station equipped with a three-element Uniform Linear Array (ULA), considering Line of Sight conditions in an open environment. The testbed, employing Ettus E312 as the transmitter and Ettus N310 as the receiver, details waveform structures and reception processes. Additionally, our study examines the performance of Angle of Arrival (AoA) estimation algorithms, such as Multiple Signal Classification (MUSIC), Estimation of Signal Parameters via Rotational Invariant Techniques (ESPRIT), and Joint Angle and Delay Estimation (JADE) ESPRIT. A MATLAB ray tracing propagation model of the environment where the measurements are conducted, has been constructed. Simulation results using this model are presented, along with the actual measurements. The obtained results affirm the effectiveness of our implementation.
\end{abstract}


\section{Introduction}


Positioning using a single node or station is an alternative way to determine the user position relying on angular and distance measurements, most suitable for the environments with limited availability of the Global Navigation Satellite System (GNSS), such as indoors or deep urban canyons. With the emergence of the mm-wave frequency range 2 (FR2) signals and large antenna arrays in 5th Generation (5G) systems, positioning using a single node has recently attracted increased research interest. A Long Term Evolution (LTE) localization testbed based on the Direction of Arrival-Time of Arrival (DoA-ToA) has been implemented by Blanco \emph{et al.} \cite{LTEtb}. Sun \emph{et al.} performed a study on 3D positioning, expanding the Multiple Signal Classification (MUSIC) algorithm to compute three parameters (azimuth angle, elevation angle and delay) and comparing the results with the Expectation-Maximization (EM) algorithm \cite{3D}. Li \emph{et al.} propose a joint Angle of Arrival (AoA) and Time of Flight (ToF) method with a single base station, utilizing Channel State Information (CSI) \cite{Li}. The MUSIC algorithm on a 4-element Uniform Linear Array (ULA) is implemented on Universal Software Radio Peripheral (USRP) nodes using LabVIEW platform by Tugrel \emph{et al.} in \cite{tugrel}. In the same sense, Rares \emph{et al.} evaluated MUSIC and Estimation of Signal Parameters via Rotational Invariance Techniques (ESPRIT) algorithms using National Instrument devices in \cite{Rares}.

 This paper focuses on the detailed modeling, simulation of real-world conditions, and experimentation of AoA estimation algorithms using Software Defined Radios (SDRs) in Uplink scenarios, utilizing the Sounding Reference Signal (SRS). The angular estimation techniques studied herein are integrated into a positioning testbed featuring a single base station. We present our comprehensive exploration of various AoA techniques in 5G through simulations, which initially informed the preliminary design of our testbed. Subsequently, we executed practical tests using real signals on the established testbed. The presented analysis, sheds light on the state-of-the-art AoA estimation algorithms and their performance metrics. The inclusion of real scenario results in conjunction with simulations has provided valuable insights. This iterative approach not only strengthens the reliability of our findings but also positions our testbed as a robust platform for assessing the practical performance of diverse 5G technologies.

The paper is organised as follows: 
Section~\ref{Methodology} presents essential information about the transmitted waveforms, the implemented channel for simulations and the utilized signal processing algorithms and methods. Section~\ref{testbed} offers an overview of the testbed, outlining its key features, detailing its components and providing a comprehensive understanding of its setup. Moving forward, Section~\ref{results} provides a summary of the simulation outcomes and the results obtained from field tests. Finally, Section~\ref{conclusions} summarizes the paper, offering concluding remarks and insights.

\section{Methodology}
\label{Methodology}
\subsection{Waveform Structure}
As proposed by the 5G standard, the SRS is used for uplink positioning. The transmitted SRS sequence is generated and mapped into the allocated subcarriers according to \cite{3GPP}. Table \ref{Configuration} describes the parameters for the numerous 5G NR waveform configurations supported by the testbed. These configurations have been identified based on different deployment scenarios (static, pedestrian, vehicular). For static and pedestrian use cases, a subcarrier spacing $\Delta f {=} 30$~kHz is considered, which is well-suited for low mobility scenarios. For the vehicular use case, the subcarrier spacing of $\Delta f {=} 60$~kHz offers more robustness to Doppler effect in high mobility scenarios, such as in vehicular environments, and to avoid inter-carrier interference (ICI). 
In the scope of this paper, only waveform configurations I, II, and III are 
analyzed.

In addition, the SRS spans 4 consecutive OFDM symbols, transmitted over the whole signal bandwidth, periodically in every slot and mapped to the physical resources according to a comb-like pattern every $K_{TC} {=} 2$ subcarriers, which provides the highest density of SRS pilots in the frequency domain. After the subcarrier allocation, the known signal is transformed in the time domain having the form:

\begin{table}[tb]
\centering
\caption{5G Waveform Configurations\vskip9pt}
\label{Waveform Configurations}
\resizebox{\textwidth}{!}{%
\footnotesize
\begin{tabular}{lcccccl}\toprule
Configuration & Numerology & Frequency Band & Subcarrier & Bandwidth \\
& & & Spacing (kHz) & (MHz)\\
\midrule
I - \hspace{0.18cm}Static, Pedestrian& $\mu$ = 1 & ISM 2.4 GHz & 30 & 20 \\ 
II - \hspace{0.1cm}Static, Pedestrian& $\mu$ = 1  & ISM 2.4 GHz & 30 & 50 \\
III - Static, Pedestrian& $\mu$ = 1 & Licensed 3.5 GHz & 30 & 20 \\ 
IV - Static, Pedestrian& $\mu$ = 1  & Licensed 3.5 GHz & 30 & 50 \\
V - \hspace{0.28cm}Vehicular& $\mu$ = 2  & Licensed 3.5 GHz & 60 & 20 \\ 
VI - \hspace{0.19cm}Vehicular& $\mu$ = 2  & Licensed 3.5 GHz & 60 & 50 \\
VII - \hspace{0.095cm}Vehicular& $\mu$ = 2 & ISM 5.8 GHz & 60 & 20 \\ 
VIII - Vehicular& $\mu$ = 2 & ISM 5.8 GHz & 60 & 50 \\
\bottomrule
\end{tabular}
}
\label{Configuration}
\end{table}
\begin{equation}
    \textbf{S(t)} = [s(t_0), s(t_1),\ldots,s(t_{{N_{\text{FFT}}\times N_{\text{OFDMsymbols}}}-1})].
\end{equation}

\subsection{Channel Model}
\label{channel}
Simulations were carried out via the MATLAB \cite{MATLAB} ray tracing propagation model \cite{raytracing}, in the field trials environment, in every frequency band that was intended to be employed (2.4~GHz, 3.5~GHz). The shooting and bouncing (SBR) method was used for the creation of the rays, with a maximum of one bounce per ray.
\begin{figure}[t]
    \centering
    \includegraphics[width=\textwidth]{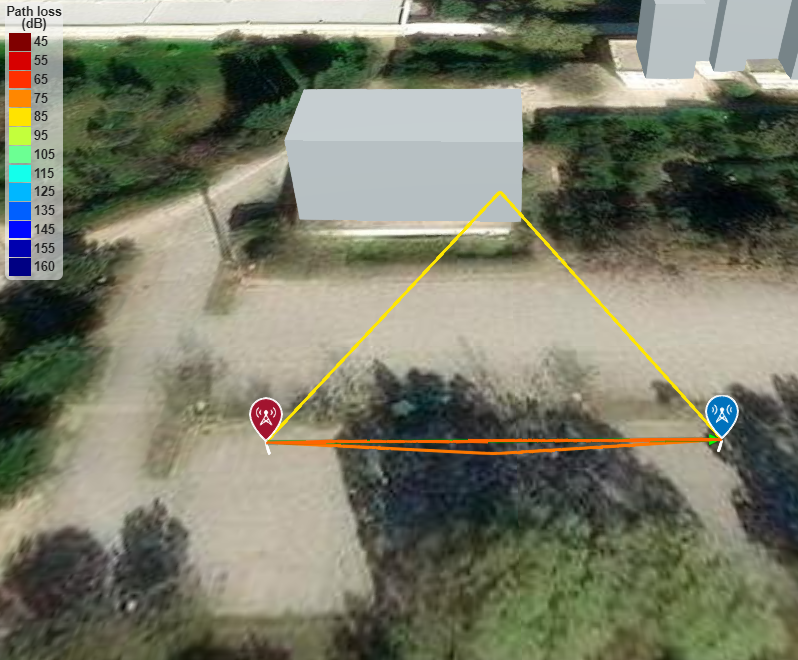}
    \caption{Propagation environment and ray creation via MATLAB ray tracing model}
    \label{raytracing}
\end{figure}

The simulation environment shown in Fig.~\ref{raytracing}, reveals a clear field, with the only notable exception being the presence of a nearby building. An extra ray is generated through reflection off the ground. According to the model, the reflection from the nearby building is not significant for distances less than 20 meters or greater than 45 meters.

\subsection{Signal Processing Algorithms}
Assuming a $M$-element ULA, $M$ copies of the transmitted signal propagate through the channel and are received, one per antenna element, having the form
\begin{equation}
    x(t) = \sum_{p=1}^{P} \mathbf{A}(\theta_p)D_p(t) s(t-\tau_p) + n(t),
\end{equation}
where $\mathbf{A}(\theta_p$) is the steering vector of the $p$-th path, 

\begin{equation}
       \mathbf{A}(\theta_p) = \begin{bmatrix}
        1\\
        \exp\left({\frac{-j2\pi f_c d \sin{\theta_p}}{c}}\right)\\
        \vdots\\
        \exp\left({\frac{-j2\pi f_c (M-1) d \sin{\theta_p}}{c}}\right)\\
    \end{bmatrix},
    \label{steering}
\end{equation}
where $f_c$ is the carrier frequency, $d$ is the antenna element spacing, $c$ is the speed of light, and $\theta_p$ is the azimuth angle of path $p$. 

\subsubsection{Timing Synchronization/Slot Detection}
Auto-correlation and cross-correlation methods have been explored for the timing synchronization of the signal, and the detection of the beginning of the slot. Since the received waveform is known at the base station, a cross-correlation method is preferred as the waveform of reference is stored/generated at the receiver side and is not subjected to noise. The offset of the received waveform in samples, compared to the original one is computed as the index $n^*$, where the largest peak of the output of the cross-correlator $c(n)$ occurs,
\begin{align}
\label{max_cross_corr}
n^* = \operatorname{argmax} c(n),
\intertext{where}
\label{cross_corr}
c(n) = \sum_{i=1}^{L_{\text{seq}} -1} s^*(i) x(n+i),
\end{align}
$L_{\text{seq}}$ denotes the length of the transmitted waveform in samples and $s^*$ denotes the conjugate of $s$.
\subsubsection{AoA Estimation Algorithms}
\label{AoA Estimation}
Three conventional Angle-of-Arrival algorithms have been studied and implemented: MUSIC \cite{MUSIC}, ESPRIT \cite{ESPRIT} and Joint Angle and Delay Estimation (JADE) ESPRIT \cite{2DESPRIT}. 

MUSIC is a super-resolution direction-finding algorithm, based on the eigenvalue decomposition of the received signal covariance matrix. The received signal $x(t)$ is transformed in the frequency domain via the Fast-Fourier Transform (FFT) operation, to obtain $\mathbf{Y}$. Considering that one subcarrier represents a single measurement, $\mathbf{Y}$ has dimensions $N_{\text{antennas}}\times N_{\text{subcarriers}}$. The covariance matrix of $\mathbf{Y}$ is 
\begin{equation}
    \textbf{R} = \text{E} [\mathbf{YY^T}].
    \label{cov matrix}
\end{equation}

As per \eqref{cov matrix}, the covariance matrix has dimensions $M \times M$. This results in the MUSIC algorithm being able to detect up to $M -1$ sources. The eigenvectors corresponding to the $D$ larger eigenvalues of the covariance matrix span the signal subspace $\mathbf{U}_\text{s} = [\mathbf{v}_1,\dots,\mathbf{v}_D]$, whereas the remaining eigenvectors span the noise subspace $\mathbf{U}_\text{n} = [\mathbf{u}_{D+1},\dots,\mathbf{u}_{M-D}]$, where $D$ denotes the number of sources.

As the covariance matrix \textbf{R} is hermitian, all its eigenvectors are orthogonal to each other, meaning that the signal subspace is orthogonal to the noise subspace. The degree of orthogonality in the MUSIC algorithm is measured by
\begin{equation}
    \text{MUSIC}_{\text{Spectrum}} = \frac{1}{\mathbf{A^H U_n U_n^H A}},
\end{equation}
where $\mathbf{A}$ is the steering vector of received signal. 

ESPRIT divides the main element array into a set of subarrays. Assuming the subarrays $\mathbf{A_1}$ and $\mathbf{A_2}$, 
it holds that
\begin{equation}
    \mathbf{A_2} =  \mathbf{A_1 \Xi},
    \label{rotation}
\end{equation}
where $\mathbf {\Xi}$ is a diagonal matrix whose main diagonal entries are $\xi_i = \exp\left({\frac{-2j \pi d \sin{\theta_i}}{\lambda}}\right)$, where $d$ is the antenna element spacing, $\theta_i$ denotes the Angle-of-Arrival $\theta$ at each antenna element  and $\lambda$ denotes the wavelength. Matrix $\mathbf{\Xi}$ applies a rotation to the matrix $\mathbf{A_1}$. Following \eqref{rotation}, ESPRIT exploits similar rotations in matrices formed by the eigenvectors of the covariance matrix of the measured data. 

After eigenvalue decomposition is performed and the signal subspace is separated from the noise subspace in a similar manner to the MUSIC algorithm, a matrix~$\mathbf{S}$ is formed,
\begin{equation}
\mathbf{S} = \mathbf{U_s}(:,1{:}D), \label{e:SU}
\end{equation}
where $\mathbf{U_s}$ is the matrix containing the eigenvectors. Notation in~\eqref{e:SU} denotes that $\mathbf{S}$ comprises the first $D$ columns of $\mathbf{U}$.

There exists a matrix $\mathbf{P}$ that contains rotational information such that the first set of eigenvectors yield the second set
\begin{equation}
    \mathbf{S_2} = \mathbf{S_1} \mathbf{P},
\end{equation}
which can be obtained via the Least Squares method, i.e.,
\begin{equation}
    \mathbf{P} = \frac{\mathbf{S}_1^* \mathbf{S}_2}{\mathbf{S}_1^* \mathbf{S}_1}.
\end{equation}

Lastly, the angle $\theta$ can be estimated in closed form, as
\begin{equation}
\theta  = \arcsin(\kappa),
\end{equation}
where
 $   \kappa  = \frac{\phi_i}{2\pi d} $
and~$\phi_i$ is the $i$-th phase angle of the total $K$ eigenvalues of~$\mathbf{P}$.

2D ESPRIT forms a Hankel matrix by stacking copies of CSI matrix $\mathbf{H}$. Similarly to 1D ESPRIT, the shift-invariant properties of the matrix are exposed. However, in this case, similar to the matrix $\mathbf{\Xi}$, a matrix $\mathbf{\Psi}$ is defined, whose main diagonal entries are $\psi_i = \exp\left({\frac{-2j \pi \tau_i}{L}}\right)$, where $L$ is the channel length measured in symbol periods. A data model given by
\begin{equation}
    \textbf{H} = \textbf{ABF}
    \label{model}
\end{equation}
is satisfied, where $\mathbf{A}$ is the Khatri-Rao product of the steering matrix with the delay matrix, $\mathbf{B}$ denotes the path attenuation and $\mathbf{F}$ is the DFT matrix with a Vandermonde structure. 

A set of selection matrices is also defined, in which $\xi_i$ and $\psi_i$ corresponding to the angles and delays, respectively, are estimated. The factor $\mathbf{F}$ in \eqref{model} ensures that a pairing between the angles and delays is satisfied. The correct pairing is carried out by a joint diagonalization procedure. To reduce complexity, all the computations can be kept in the real domain as described in \cite{UnitaryESPRIT}.

\subsubsection{SINR Computation}
\label{SINR}
A crucial metric in assessing performance is the Signal-to-Interference-plus-Noise Ratio (SINR) computation. As previously articulated, our signal transmission employs a comb-like pattern every
$K_{TC} {=} 2$ subcarriers, wherein every alternate subcarrier remains unoccupied. Consequently, we compute the power associated with these vacant subcarriers, constituting the noise component. By subtracting this noise power from the total power of the utilized subcarriers, we ascertain the signal power. Subsequently, the SINR for each time slot is computed as 
\begin{equation}
    \text{SINR} = 10 \log_{10}\left(\frac{\text{P}_{\text{utilized subcarriers}} - \text{P}_{\text{empty subcarriers}}}{\text{P}_{\text{empty subcarriers}}}\right).
\end{equation}

\section{Testbed Description}
\label{testbed}
 This testbed employs the transmission of representative 5G waveforms through SDRs, with the base station featuring a three-element ULA. A 5G uplink waveform containing a number of known SRS sequences depending on the bandwidth is generated and transmitted by the user in the desired frequency. The user is responsible for generating, mapping and transmitting the SRS sequences while the receiver processes the received signal and performs timing synchronization and AoA estimation.

\begin{figure}
    \centering
    \includegraphics[width=\textwidth]{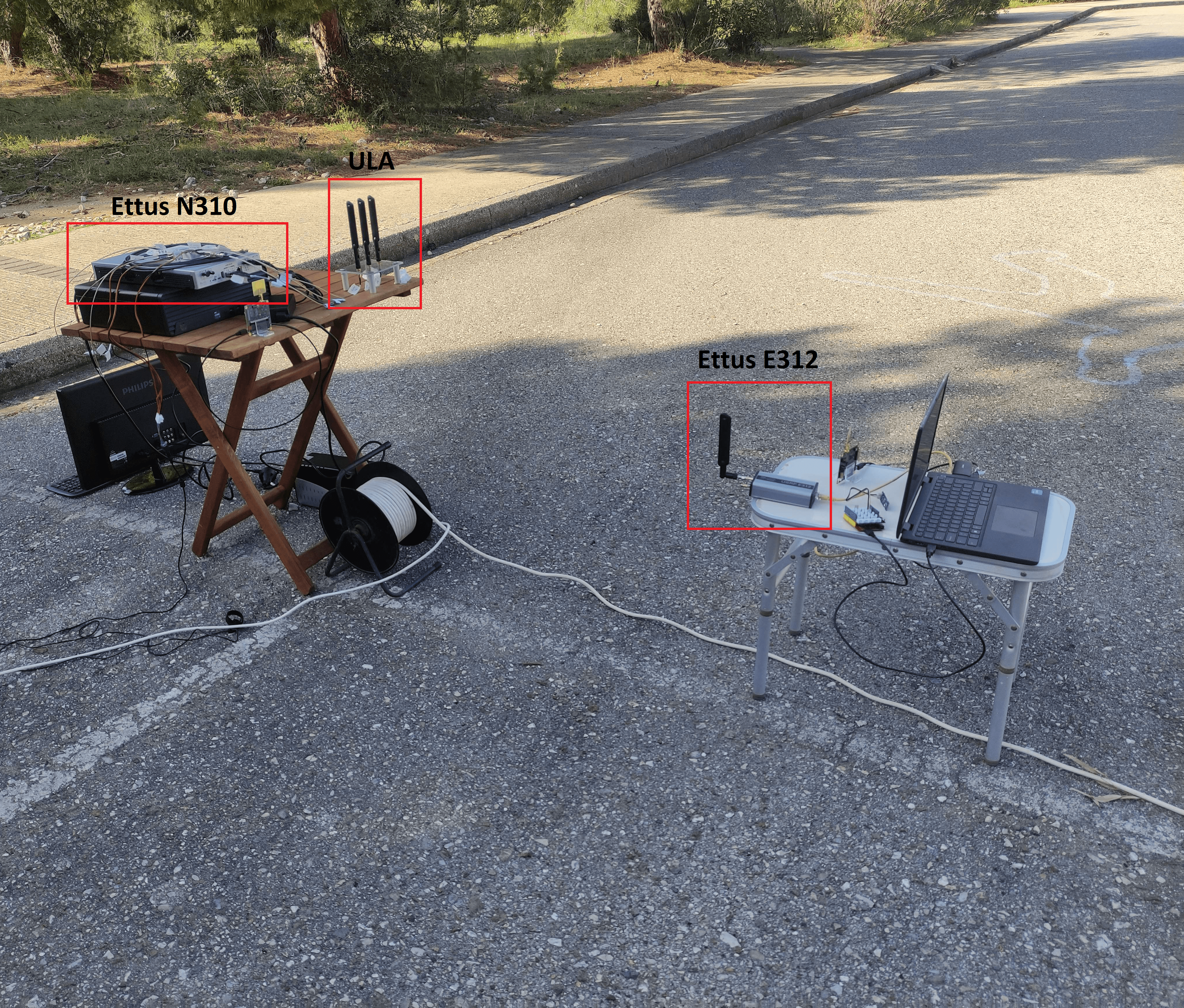}
    \caption{User and Base Station setup during field trials}
    \label{Setup}
\end{figure}
\subsection{Testbed Equipment}
\begin{figure}
    \centering
    \includegraphics[width=\textwidth]{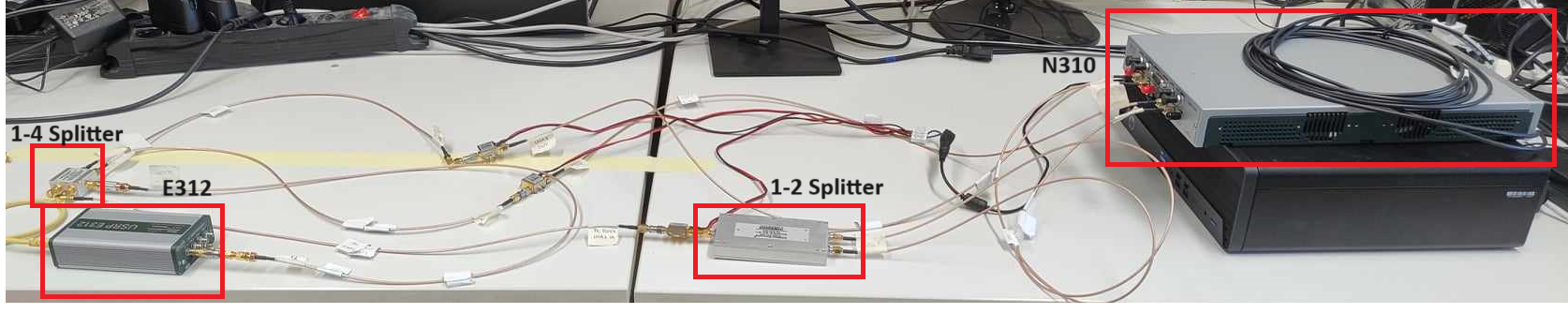}
    \caption{User and Base Station setup during the calibration procedure}
    \label{singpos_cal_eq}
\end{figure}

The testbed setup utilizes an Ettus E312 as the transmitter and an Ettus N310 as the receiver, presented in Fig.~\ref{Setup}. Although the N310 has four RX channels, only three are used for AoA estimation. This decision is driven by the N310's architecture, which includes two daughterboards, each with a pair of RX channels. All four channels of the N310 are originally misaligned in phase, necessitating a phase offset compensation procedure. At first, phase offset compensation is performed independently for the channel pairs within the N310 by feeding a tone signal to all four channels using a 1-4 splitter. The phase difference between the two channels of each pair is then computed by cross-correlating the received signals. These computed values are stored and applied during signal processing to correct phase offsets, ensuring phase alignment within each pair of channels.
Moreover, because the N310's two daughterboards use different Local Oscillators (LOs) for their respective RX channel pairs, the N310 cannot inherently align these pairs as per \cite{Xhafa23}. This results in random phase variations between runs. To address these phase offsets from different LO initializations, a real-time calibration process is introduced. For this procedure, as the testbed normally operates, a common signal is injected into one channel of each pair via a 1-2 splitter, allowing the differential phase due to the different LOs to be measured, but limiting the available channels for AoA estimation to three. This inherent phase difference is then compensated in real-time, ensuring overall phase alignment. The testbed setup for the initial phase offset compensation procedure is depicted in Fig.~\ref{singpos_cal_eq}. Furthermore, dedicated software has been developed to control and manage the testbed during experimentation. This software facilitates seamless coordination, ensuring the overall optimization of the experimental setup.

\subsection{Testbed Signal Processing}
Given the potentially impractical size of the IQ samples file, both in terms of storage and processing efficiency, a snapshot technique has been implemented. Recognizing the necessity of obtaining one angular estimation per second, this method ensures that only a fraction of milliseconds for each second of the captured signal is retained on the host PCs, thereby significantly diminishing the overall file size. In the context of 5G numerologies 1 and 2, relevant to our work, one slot corresponds to 0.5 ms and 0.25 ms, respectively. Consequently, capturing 1 ms of signal is considered sufficient in all scenarios, as it aligns with the presence of a whole slot at all times.

The signal processing scheme for one angular measurement each second, is described in Algorithm~\ref{Signal Processing}.
\begin{algorithm}[t]
    \caption{AoA Estimation with Ettus N310}
    \label{Signal Processing}
    \begin{algorithmic}[1]
        \WHILE{remaining size of the IQ Samples is greater than or equal to the size of a snapshot}
        \STATE Align the phase of the snapshot for the individual channels of the two pairs, using values computed in offline calibration.
        \STATE Initialize a pointer at the first IQ sample. Load IQ samples corresponding to one snapshot.
        \WHILE{remaining size of snapshot is greater than or equal to twice the size of a slot}
            \STATE Load IQ samples equivalent to two slots.
            \STATE Determine the start of the 5G slot by cross-correlating loaded IQ samples with the known waveform using \eqref{max_cross_corr}, \eqref{cross_corr}.
            \STATE Align the phase of the two channel pairs by computing the phase difference of the common signal.
            \STATE Transform the received signal in the frequency domain by removing the cyclic prefix and performing FFT. Form a grid for each antenna, with size $N_{\text{OFDM Symbols Per Slot}} \times N_{\text{Subcarriers}}$.
            \STATE Extract the first 4 OFDM symbols of the slot that contain the SRS pilots.
            \STATE Estimate Signal-to-Interference-plus-Noise Ratio (SINR) as described in Section~\ref{SINR}.
            \STATE Perform AoA estimation as outlined in Section~\ref{AoA Estimation}.
        \ENDWHILE
        \STATE Remove outliers that deviate more than three scaled Median Absolute Deviations (MAD) from the median of the data.
        \STATE Average the remaining SINR and angular estimations of the snapshot. Increment the pointer by the number of IQ samples corresponding to one snapshot.
    \ENDWHILE
    \end{algorithmic}
\end{algorithm}

\section{Results and Discussion}
\label{results}
\subsection{Simulations}
\label{Simulation Results}
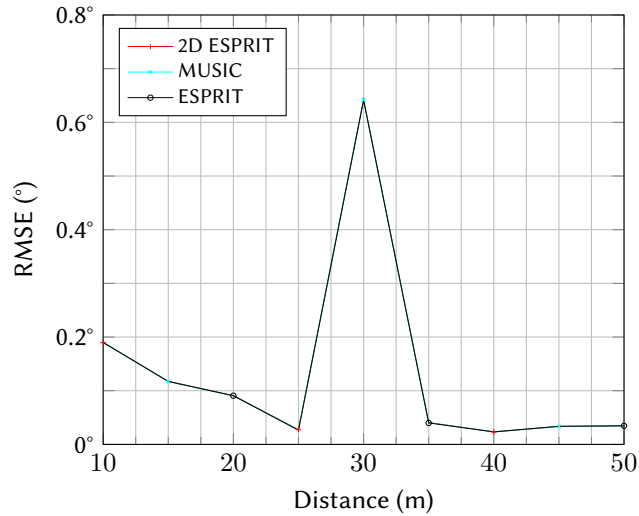
\begin{figure}[t]
\centering
\begin{tikzpicture}
\centering
\begin{axis}[legend pos=north west, 
        legend cell align={left},
        legend style={nodes={scale=0.8, transform shape}},
        xlabel=Distance (m), ylabel=RMSE ($\degree$), xmin=10, xmax=50, ymin=0, 
        ytick={0,0.1,...,0.8},
        yticklabels={0\degree{},,0.2\degree{},,0.4\degree{},,0.6\degree{},,0.8\degree{}},
        ymax=0.8,
        grid=both,
        minor x tick num=3]

    \addplot[red,mark=+,mark size=1pt,
    mark repeat=3, mark phase=1, very thin] table[x index = 0, y index = 1 ]{Data/ray_tracing_parking_sims_2_4_v2.tex};
    \addplot[cyan,mark=x,mark size=1pt,
    mark repeat=3, mark phase=2, very thin] table[x index = 0, y index = 2 ]{Data/ray_tracing_parking_sims_2_4_v2.tex};
    \addplot[black,mark=o,mark size=1pt,
    mark repeat=3, mark phase=3, very thin] table[x index = 0, y index = 3 ]{Data/ray_tracing_parking_sims_2_4_v2.tex};
   
    \addlegendentry{2D ESPRIT}
    \addlegendentry{MUSIC}
    \addlegendentry{ESPRIT}

\end{axis}
\end{tikzpicture}
\caption{RMSE in degrees for reception with three antennas, simulated with ray tracing model for the field trials area, at 2.4 GHz, with LOS at 0$\degree{}$.}
\label{2_4_sims}
\end{figure}

\begin{figure}[t]
\centering
\begin{tikzpicture}
\begin{axis}[legend pos=north west, 
        legend cell align={left},
        legend style={nodes={scale=0.8, transform shape}},
        xlabel=Distance (m), ylabel=RMSE ($\degree$), xmin=10, xmax=50, ymin=0, 
        ytick={0,0.1,...,0.8},
        yticklabels={0\degree{},,0.2\degree{},,0.4\degree{},,0.6\degree{},,0.8\degree{}},
        ymax=0.8,
        grid=both,
        minor x tick num=3]

    \addplot[red,mark=+,mark size=1pt,
    mark repeat=3, mark phase=1, very thin] table[x index = 0, y index = 1 ]{Data/ray_tracing_parking_sims_3_5_v2.tex};
    \addplot[cyan,mark=x,mark size=1pt,
    mark repeat=3, mark phase=2, very thin] table[x index = 0, y index = 2 ]{Data/ray_tracing_parking_sims_3_5_v2.tex};
    \addplot[black,mark=o,mark size=1pt,
    mark repeat=3, mark phase=3, very thin] table[x index = 0, y index = 3 ]{Data/ray_tracing_parking_sims_3_5_v2.tex};
   
    \addlegendentry{2D ESPRIT}
    \addlegendentry{MUSIC}
    \addlegendentry{ESPRIT}

\end{axis}
\end{tikzpicture}
\caption{RMSE in degrees for reception with three antennas, simulated utilizing a ray tracing model for the field trials area, at 3.5~GHz, with LOS at 0$\degree$.}
\label{3_5_sims}
\end{figure}
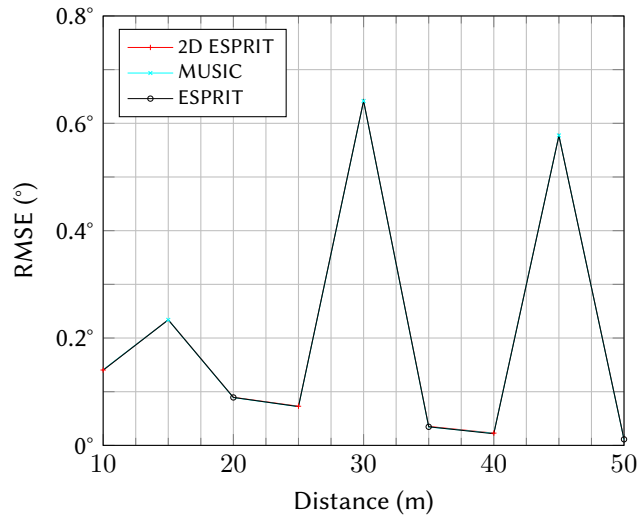

Prior to conducting field trials, we utilized the MATLAB ray tracing propagation model to simulate the performance of the three mentioned algorithms in the designated field environment, as described in Section~\ref{channel}.  Simulations were performed in the dedicated frequency bands (2.4~GHz, 3.5~GHz), for an AoA of 0$\degree$, using the corresponding 5G signals. Initial tests measuring received power were undertaken, and Additive White Gaussian Noise (AWGN) was introduced in the simulations to replicate real Signal to Noise Ratio (SNR) conditions. As the ray tracing tool offers a deterministic approach regarding the propagation channel, Monte-Carlo simulations of 200 measurements per distance for the given SNR values. Furthermore, The simulation analysis assumes perfect antenna calibration. In reality, this is not the case as antenna calibration errors decrease the accuracy of the angular estimation. 

The simulation results, illustrated in Figs.~\ref{2_4_sims} and \ref{3_5_sims}, indicate that, under conditions of short distances (below 20 meters) with a clear Line of Sight (LOS) path and only ground reflections, the algorithms exhibit more stable performance. This stability contrasts with distances involving reflections from the nearby building, as elaborated in Section~\ref{channel}. All three algorithms exhibit similar performance across both frequency bands, with occasional spikes in effectiveness observed in the presence of reflections.

\subsection{Field Tests}
Preliminary field trials were conducted on the University of Patras campus to validate the operational capabilities of the testbed and evaluate the efficacy of Super-Resolution AoA estimation algorithms with 5G signals in real scenarios. The positioning of the transmitter (Ettus E312) and the receiver (Ettus N310) adhered to the parameters established in the simulations outlined in Section~\ref{Simulation Results}. As detailed in the aforementioned section, simulations highlighted a significant impact on algorithm performance due to a robust reflection from a nearby building.

\begin{table}[tb]
\centering
\caption{Tests Overview\vskip9pt}
\label{Field Trial Configurations}
\resizebox{\textwidth}{!}{%
\footnotesize
\begin{tabular}{lccccccl}\toprule
Configuration & Frequency Band & Angle of Arrival & Test Duration (s) & Distance (m) & Snapshot Length\\
\midrule
I & ISM 2.4 GHz & 0$\degree{}$ & 60 & 15 & 3 ms\\
II & ISM 2.4 GHz & 0$\degree{}$ & 60 & 15 & 3 ms\\
III & Licensed 3.5 GHz & 0$\degree{}$ & 60 & 15 & 3 ms\\
III & Licensed 3.5 GHz & 5$\degree{}$ & 60 & 15 & 3 ms\\
III & Licensed 3.5 GHz & 10$\degree{}$ & 60 & 15 & 3 ms\\
III & Licensed 3.5 GHz & 15$\degree{}$ & 60 & 15 & 3 ms\\
III & Licensed 3.5 GHz & 20$\degree{}$ & 60 & 15 & 3 ms\\
III & Licensed 3.5 GHz & 25$\degree{}$ & 60 & 15 & 3 ms\\
III & Licensed 3.5 GHz & 45$\degree{}$ & 60 & 15 & 3 ms\\
I & ISM 2.4 GHz & 0$\degree{}$ & 60 & 50 & 3 ms\\
II & ISM 2.4 GHz & 0$\degree{}$ & 60 & 50 & 3 ms\\
\bottomrule
\end{tabular}
}
\label{tests}
\end{table}

As the scope of this work targeted an open-field setting, two series of tests were undertaken to minimize the impact of multipath: one at a close proximity of 15 meters and another at a greater distance of 50 meters. This first set of tests was conducted in both the Industrial Scientific and Medical (ISM) 2.4 GHz band and the Licensed 3.5 GHz band, using configurations~I,~II and~III from Table~\ref{Waveform Configurations}. The second set of tests was conducted in the ISM 2.4 GHz band, using configurations~I and II from Table~\ref{Waveform Configurations}. Table~\ref{tests} outlines the conducted tests. The angle of arrival for the 2.4 GHz band tests was fixed at 0$\degree$, while tests for the 3.5 GHz band were performed across the range of 0$\degree$ to 25$\degree$ with a step of 5$\degree$. An additional test was executed at 45$\degree$. In all tests, a snapshot length of 3 ms was selected. Since all three configurations (I, II, III) use numerology $\mu = 1$, each snapshot contains 5 slots, resulting in 5 AoA estimations per snapshot, and therefore per second.

\begin{figure}[!h]
\centering
\begin{tikzpicture}
\begin{axis}[legend pos=north east, 
        legend cell align={left},
        legend style={nodes={scale=0.45, transform shape}},
        xlabel=Time (s), ylabel=Angle ($\degree$), xmin=1, xmax=60, ymin=-5, 
        ytick={0,5,10,15,20,25,45},
        yticklabels={0\degree{},
        5\degree{},
        10\degree{},
        15\degree{},
        20\degree{},
        25\degree{},
        45\degree{}},
        ymax=55,
        grid=both,
        minor x tick num=3]
   
    \addplot[red,mark=+,mark size=1pt,
    mark repeat=3, mark phase=1, very thin] table[x index = 0, y index = 1 ]{Data/sigid3.tex};
    \addplot[cyan,mark=x,mark size=1pt,mark repeat=3, mark phase=2, very thin] table[x index = 0, y index = 2 ]{Data/sigid3.tex};
    \addplot[black,mark=o, mark size=1pt,mark repeat=3, mark phase=3,very thin] table[x index = 0, y index = 3 ]{Data/sigid3.tex};
    \addplot[red,mark=+,mark size=1pt,
    mark repeat=3, mark phase=1, very thin] table[x index = 0, y index = 4 ]{Data/sigid3.tex}; 
    \addplot[cyan,mark=x,mark size=1pt,mark repeat=3, mark phase=2, very thin] table[x index = 0, y index = 5 ]{Data/sigid3.tex};
    \addplot[black,mark=o, mark size=1pt,mark repeat=3, mark phase=3,very thin] table[x index = 0, y index = 6 ]{Data/sigid3.tex};
    \addplot[red,mark=+,mark size=1pt,
    mark repeat=3, mark phase=1, very thin]  table[x index = 0, y index = 7 ]{Data/sigid3.tex};
    \addplot[cyan,mark=x,mark size=1pt,mark repeat=3, mark phase=2, very thin] table[x index = 0, y index = 8 ]{Data/sigid3.tex};
    \addplot[black,mark=o, mark size=1pt,mark repeat=3, mark phase=3,very thin] table[x index = 0, y index = 9 ]{Data/sigid3.tex};
    \addplot[red,mark=+,mark size=1pt,
    mark repeat=3, mark phase=1, very thin] table[x index = 0, y index = 10 ]{Data/sigid3.tex};
    \addplot[cyan,mark=x,mark size=1pt,mark repeat=3, mark phase=2, very thin] table[x index = 0, y index = 11 ]{Data/sigid3.tex};
    \addplot[black,mark=o, mark size=1pt,mark repeat=3, mark phase=3,very thin] table[x index = 0, y index = 12 ]{Data/sigid3.tex};
    \addplot[red,mark=+,mark size=1pt,
    mark repeat=3, mark phase=1, very thin] table[x index = 0, y index = 13 ]{Data/sigid3.tex};
    \addplot[cyan,mark=x,mark size=1pt,mark repeat=3, mark phase=2, very thin] table[x index = 0, y index = 14 ]{Data/sigid3.tex};
    \addplot[black,mark=o, mark size=1pt,mark repeat=3, mark phase=3,very thin] table[x index = 0, y index = 15 ]{Data/sigid3.tex};
    \addplot[red,mark=+,mark size=1pt,
    mark repeat=3, mark phase=1, very thin] table[x index = 0, y index = 16 ]{Data/sigid3.tex};
    \addplot[cyan,mark=x,mark size=1pt,mark repeat=3, mark phase=2, very thin] table[x index = 0, y index = 17 ]{Data/sigid3.tex};
    \addplot[black,mark=o, mark size=1pt,mark repeat=3, mark phase=3,very thin] table[x index = 0, y index = 18 ]{Data/sigid3.tex};
    \addplot[red,mark=+,mark size=1pt,
    mark repeat=3, mark phase=1, very thin] table[x index = 0, y index = 19 ]{Data/sigid3.tex};
    \addplot[cyan,mark=x,mark size=1pt,mark repeat=3, mark phase=2, very thin] table[x index = 0, y index = 20 ]{Data/sigid3.tex};
    \addplot[black,mark=o, mark size=1pt,mark repeat=3, mark phase=3,very thin] table[x index = 0, y index = 21 ]{Data/sigid3.tex};
   
    \addlegendentry{2D ESPRIT}
    \addlegendentry{MUSIC}
    \addlegendentry{ESPRIT}

\end{axis}
\end{tikzpicture}
\caption{Performance evaluation of super-resolution algorithms in Licensed 3.5~GHz band for static tests for different Angles of Arrival at 15~m distance}
\label{3_5}
\end{figure}
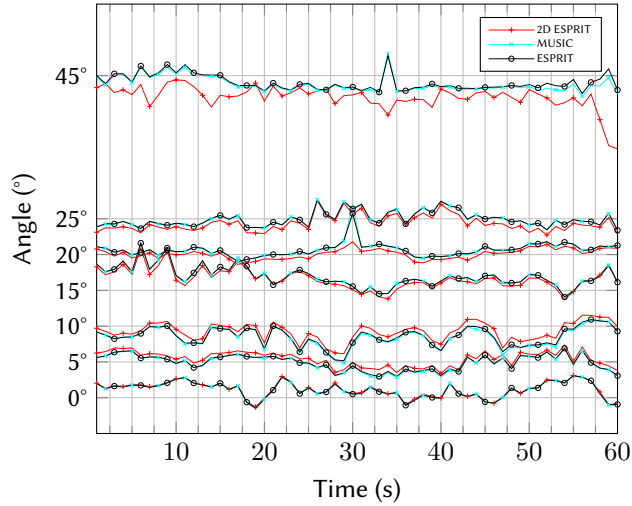

The outcomes of the static tests at the 3.5 GHz band, considering various angles of arrival at a distance of 15~meters over a duration of 60~seconds, are illustrated in Fig.~\ref{3_5}. Evaluation of the angle of arrival resolution algorithms consistently demonstrates similar performance across all scenarios, affirming the results obtained from the simulations. Minimal fluctuations are observed, with particular notability in the cases of the MUSIC and ESPRIT algorithms.

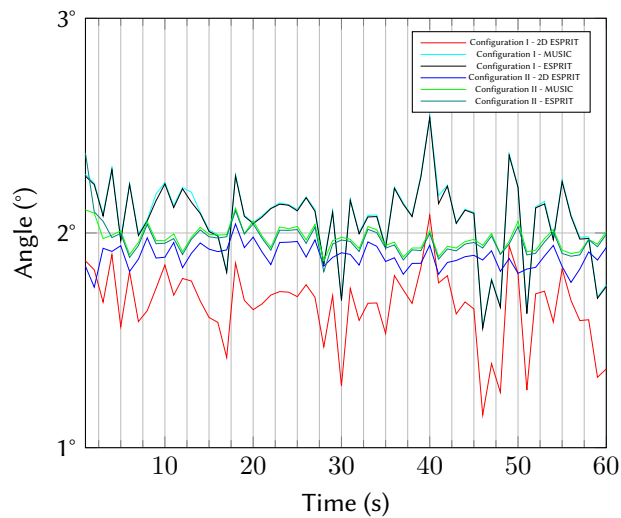
\begin{figure}[!h]
\centering
   \begin{tikzpicture}[scale=1, transform shape]
    \begin{axis}[
    legend pos= north east,
    legend style={nodes={scale=0.35, transform shape}}, 
        legend image post style={mark=none},xlabel=Time (s), ylabel=Angle ($\degree$), xmin=1, xmax=60, ymin=1, ymax=3,
        ytick={1,2,3},
        yticklabels={
        1\degree{},
        2\degree{},
        3\degree{}},
        ymax=3,
        grid=both,
        minor x tick num=3]
    \addplot[red,mark=none,mark size=2pt,
    mark repeat=3, mark phase=1, very thin] table[x index = 0, y index = 1 ]{Data/aoa_15m.tex};
    \addplot[cyan,mark=none,mark size=1pt,mark repeat=3, mark phase=2, very thin] table[x index = 0, y index = 2 ]{Data/aoa_15m.tex};
    \addplot[black,mark=none, mark size=1pt,mark repeat=3, mark phase=3,very thin] table[x index = 0, y index = 3 ]{Data/aoa_15m.tex};
    \addplot[blue,mark=none,mark size=1pt,
    mark repeat=3, mark phase=1, very thin] table[x index = 0, y index = 4 ]{Data/aoa_15m.tex}; 
    \addplot[green,mark=none,mark size=1pt,mark repeat=3, mark phase=2, very thin] table[x index = 0, y index = 5 ]{Data/aoa_15m.tex};
    \addplot[teal,mark=none, mark size=1pt,mark repeat=3, mark phase=3,very thin] table[x index = 0, y index = 6 ]{Data/aoa_15m.tex};

    \addlegendentry{Configuration I - 2D ESPRIT}
    \addlegendentry{Configuration I - MUSIC}
    \addlegendentry{Configuration I - ESPRIT}
    \addlegendentry{Configuration II - 2D ESPRIT}
    \addlegendentry{Configuration II - MUSIC}
    \addlegendentry{Configuration II - ESPRIT}

\end{axis}
\end{tikzpicture}
\caption{Angle of Arrival Estimation at 15~m distance in ISM 2.4~GHz band, using Configurations I and II, with LOS at 0$\degree$}
\label{15m}
\end{figure}

\begin{figure}[!h]
\centering
   \begin{tikzpicture}[scale=1, transform shape]
    \begin{axis}[
    legend pos= north east,
    legend style={nodes={scale=0.35, transform shape}}, 
        legend image post style={mark=none},xlabel=Time (s), ylabel=Angle ($\degree$), xmin=1, xmax=60, ymin=-2, 
        ytick={0,-1,-2},
        yticklabels={0\degree{},
        $-$1\degree{},
        $-$2\degree{}},
        ymax=0,
        grid=both,
        minor x tick num=3]
    \addplot[red,mark=none,mark size=2pt,
    mark repeat=3, mark phase=1, very thin] table[x index = 0, y index = 1 ]{Data/aoa_50m.tex};
    \addplot[cyan,mark=none,mark size=1pt,mark repeat=3, mark phase=2, very thin] table[x index = 0, y index = 2 ]{Data/aoa_50m.tex};
    \addplot[black,mark=none, mark size=1pt,mark repeat=3, mark phase=3,very thin] table[x index = 0, y index = 3 ]{Data/aoa_50m.tex};
    \addplot[blue,mark=none,mark size=1pt,
    mark repeat=3, mark phase=1, very thin] table[x index = 0, y index = 4 ]{Data/aoa_50m.tex}; 
    \addplot[green,mark=none,mark size=1pt,mark repeat=3, mark phase=2, very thin] table[x index = 0, y index = 5 ]{Data/aoa_50m.tex};
    \addplot[teal,mark=none, mark size=1pt,mark repeat=3, mark phase=3,very thin] table[x index = 0, y index = 6 ]{Data/aoa_50m.tex};

    \addlegendentry{Configuration I - 2D ESPRIT}
    \addlegendentry{Configuration I - MUSIC}
    \addlegendentry{Configuration I - ESPRIT}
    \addlegendentry{Configuration II - 2D ESPRIT}
    \addlegendentry{Configuration II - MUSIC}
    \addlegendentry{Configuration II - ESPRIT}

\end{axis}
\end{tikzpicture}
\caption{Angle of Arrival Estimation at 50~m distance in ISM 2.4~GHz band, using Configurations I and II, with LOS at 0$\degree$}
\label{50m}
\end{figure}
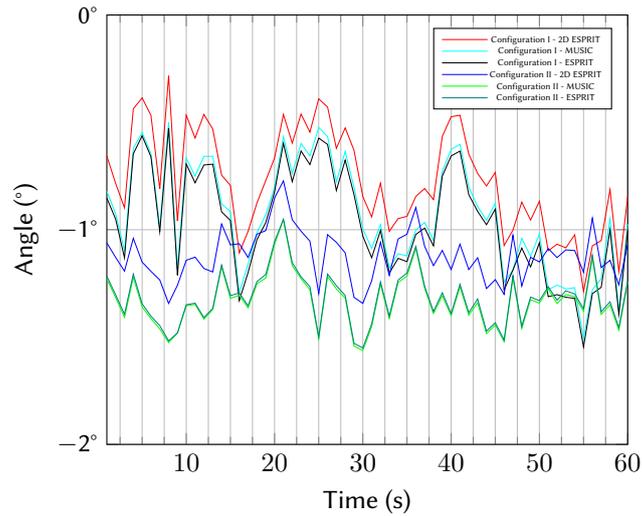

Likewise, Figs.~\ref{15m} and \ref{50m} illustrate the results of static tests conducted at the ISM band in 2.4 GHz, where the angle of arrival was fixed at 0$\degree$, spanning distances of 15 meters and 50 meters respectively. Once again, the performance of the algorithms exhibits a notable similarity, particularly when contrasted with the overall fluctuations observed in the measurements.

In evaluating the overall performance of the testbed, it is crucial to acknowledge the complexity of precisely setting the desired angle of arrival. Despite using equipment to align the transmitter with the receiver in terms of angles, height, and floor tilt, minor discrepancies may arise due to potential human error. With that said, the obtained results closely align with the desired outcome in the majority of cases. Across various scenarios, we observe an accuracy of less than 2$\degree$ of error, accompanied by consistent results throughout the entire test duration. It is noteworthy that certain significant fluctuations observed in the 3.5 GHz band test, particularly at 0$\degree$, 10$\degree$, and 15$\degree$ angle of arrival, can be attributed to small channel fluctuations and potential imperfections in the equipment. Furthermore, this particular frequency band is susceptible to large amounts of interference due to the utilization of the spectrum by the mobile providers. In conclusion, while the discrepancy between the simulation results and actual measurements may seem significant, it is crucial to differentiate the simulation environment and models from real-world conditions. The obtained results, overcoming factors such as interference, antenna array imperfections, and equipment limitations, when also combined with the real channel, highlight the robust performance of the testbed.

\section{Conclusions}
\label{conclusions}
In conclusion, this study provides a comprehensive evaluation of super-resolution algorithms in 5G uplink scenarios through a combination of simulations and real experiments, within the context of developing a positioning testbed. Field trials were emulated through simulations using MATLAB ray tracing propagation model with 5G SRS signals across various distances. Real experiments utilized Ettus E312 as the user and Ettus N310 as the base station, equipped with a three-element ULA. Calibration of N310 channels, compensating for phase offsets, was performed before signal processing. A snapshot technique for signal reception was implemented to reduce the size of received IQ sample files and processing speed. Static tests conducted at 2.4 GHz and 3.5 GHz bands demonstrated comparable performance among all evaluated AoA algorithms. Despite the preliminary nature of these tests, our testbed exhibited commendable performance, delivering stability and accuracy in its results. 

In addition to the findings presented in this study, it is noteworthy that our testbed serves as an ongoing platform for further investigations. The current work involves continuous measurements and additional experiments, particularly expanding into the 5.8 GHz band, leveraging the capabilities of the developed testbed. This sustained effort aims to enhance our understanding of 5G positioning technologies in real-world scenarios, contributing to the refinement and expansion of practical applications.

\section{Acknowledgements}
The undertaken efforts were conducted within the framework of the Single Node Positioning Testbed (SINGPOS) project funded by the European Space Agency (ESA).

\bibliography{icl_gnss}

\end{document}